\documentclass{article}

\usepackage{graphicx}
\usepackage[nonatbib, final]{neurips_2021_ml4ps}

\usepackage[utf8]{inputenc} 
\usepackage[T1]{fontenc}    
\usepackage{hyperref}       
\usepackage{url}            
\usepackage{booktabs}       
\usepackage{amsfonts, amsmath}       
\usepackage{nicefrac}       
\usepackage{microtype}      
\usepackage{xcolor} 
\usepackage{adjustbox}
\title{Learning Full Configuration Interaction Electron Correlations with Deep Learning}

\author{
Hector H. Corzo\thanks{ Center for Chemical Computation and Theory, University of California, Merced, CA.} \\
    \texttt{hhcorzo@gmail.com}
    \And    
    Arijit Sehanobish\thanks{Work done as a postdoc at Yale University. Currently at Covera Health, NYC}  \thanks{Equal Contribution} \\
    \texttt{arijit.sehanobish1@gmail.com}
    \And    
    Onur Kara\thanks{Hindsight Technology Solutions}\\
    \texttt{okara83@gmail.com}
    }

\begin{document}

\maketitle

\begin{abstract}
In this report, we present a deep learning framework termed the Electron Correlation Potential Neural Network (eCPNN) that can learn succinct and compact potential functions. These functions can effectively describe the complex instantaneous spatial correlations among electrons in many--electron atoms. The eCPNN was trained in an unsupervised manner with limited information from Full Configuration Interaction (FCI) one--electron density functions within predefined limits of accuracy. Using the effective correlation potential functions generated by eCPNN, we can predict the total energies of each of the studied atomic systems with a remarkable accuracy when compared to FCI energies.  

\end{abstract}
\section{Introduction}~\label{intro}
Applications of machine learning (ML) are ubiquitous in many branches of the natural sciences. In particular, deep neural networks (NN) are revolutionizing the approach in which problems in  molecular sciences are explored. ML techniques offer an attractive alternative for reducing the computational time, hardware demands, and the complexity needed for many--electron atom and molecular simulations.
A multitude of propositions using ML for the treatment of such quantum systems exist. NN frameworks have been constructed for atomistic simulations\cite{Dral2021,behler2011atom}, end-to-end learning of the forward and inverse Schr\"odinger equation\cite{QPNN,Manzhos2020}, molecular dynamics simulations\cite{Zhang2018,NoeSci,yang2019molecular,Manzhos2020}, and the direct representation of highly correlated wave functions \cite{CarleoSci,Pfau2020,ma2020deep}.
Such studies inspired the design of an unsupervised learning model capable of defining effective functions that can describe many--electron systems. By using limited information obtained from the electronic probability density, a NN is constructed that is able to learn an effective Hamiltonian which describes the electronic correlation effects relevant for many--electron atoms. This NN approach for describing many--electron systems has been developed on the intrinsic conceptual formalism of the configuration interaction (CI) method for solving non-relativistic many--electron correlated systems within the Born–Oppenheimer approximation. High fidelity CI data was used to directly train the NN in an unsupervised manner. 

This work focuses on three main points, \textbf{(a)} the introduction of a NN model for many-electron atoms defined in terms of FCI concepts and  high fidelity data, \textbf{(b)} the approximation of observables such as total energy for atomic system  and \textbf{(c)}
the agreement of our learned potential with the virial theorem, which although originally formulated for classical systems, has been shown to hold for quantum mechanical systems as well~\cite{lowdin1959scaling,marc1985virial,Fock1930}.
\section{Correlated Electronic Densities for Atomic Systems}~\label{FCI}
In  atomic and molecular theory, numerically exact solutions to the electronic time-independent, non-relativistic  Schr\"{o}dinger equation for many--electron systems is given by the configuration interaction matrix-eigenvalue equation~\cite{shavitt1977method,bunge2018present,barr1970nature}, 
\begin{equation}\label{eq:2}
    \textbf{HC}_{\mu}=E^{FCI}_{\mu}\textbf{C}_{\mu},
\end{equation}
where \textbf{H} is the representation of the Hamiltonian $\hat{\rm H}$ in terms of Slater determinants $D^{abc\dots}_{ijk\dots}$, ${C}_{\mu}$ are known as the variational coefficients,  and $E^{FCI}_{\mu}$ is the total energy of the system within an infinitely complete basis set. Beyond one electron systems, the operator $\hat{\rm H}$ is not separable, thus only approximate solutions  for equation \ref{eq:2} exist. If a complete set of orthonormal one-electron functions of space-spin coordinates $\chi=(\Vec{r},\xi)$ is defined, every normalizable antisymmetric wave-function can be expressed as a linear combination of $D^{abc\dots}_{ijk\dots}$. Then, the FCI wave-function  written as an expansion of variational coefficients ($c^{abc\dots}_{ijk\dots}$), in its cluster form, reads
\begin{equation}\label{eq:Cluster}
    \Psi^{FCI}= D_{0}c_{0}+\sum_{i}\sum_{a}D^{a}_{i}c^{a}_{i}+ \sum_{i<j}\sum_{a<b}D^{ab}_{ij}c^{ab}_{ij}+ \sum_{i<j<k}\sum_{a<b<c}D^{abc}_{ijk}c^{abc}_{ijk} + \cdots,
\end{equation}
where the indices $i,j,k$ denote occupied (Occ) spin-orbitals or Dirac bi-spinors, whereas the $a,b,c$ indices represent the correspond unoccupied or virtual (Virt) one-electron functions. In the $\Psi^{FCI}$, the CI space is expanded according to electron substitutions or \textit{excitations} levels from Occ to Virt orbitals defining different configurations. In this respect, non--relativistic selected configuration interaction (SCI) wave--functions, $\Psi^{SCI}$ ~\cite{bunge2018present,bunge2005cracking,bunge2006select}  provide compact wave-functions to define reliable and stable electronic densities for many--electron systems over a wide range of values of the radial coordinate ($\Vec{r}$). By definition, $\Psi^{SCI}$ yields equivalent energies  as $\Psi^{FCI}$ in a more compact representation where only the most important configurations within a predefined limits of accuracy are computed~\cite{bunge2005cracking,bunge2018present}. $\Psi^{SCI}$ are defined by a finite expansion  of configuration state functions (CSFs)  expressed as linear combination of Slater determinants, 
\begin{equation}
\Psi^{SCI}=\sum_{K,p} \phi_{K}^{p}a_{Kp},
\end{equation} 
where $\phi_{K}^{p}a_{Kp}$ represents the CSFs with  $K$ configurations and $p$ degenerate elements.
 After unitary transformation and diagonalization, the electronic probability density  of $\Psi^{SCI}$ may be expressed in terms of  natural radial orbitals, $\chi_{il}(r)$, and occupation numbers $n_{il}$, 
\begin{equation}
\rho(r)=\sum_{i}f(i,l,ms)n_{il}\chi^{2}_{il}(r),
\end{equation}
where $f(i,l,ms)$ is the $LS$  symmetry factors of the CSFs. 

\begin{figure}
\includegraphics[width=\linewidth]{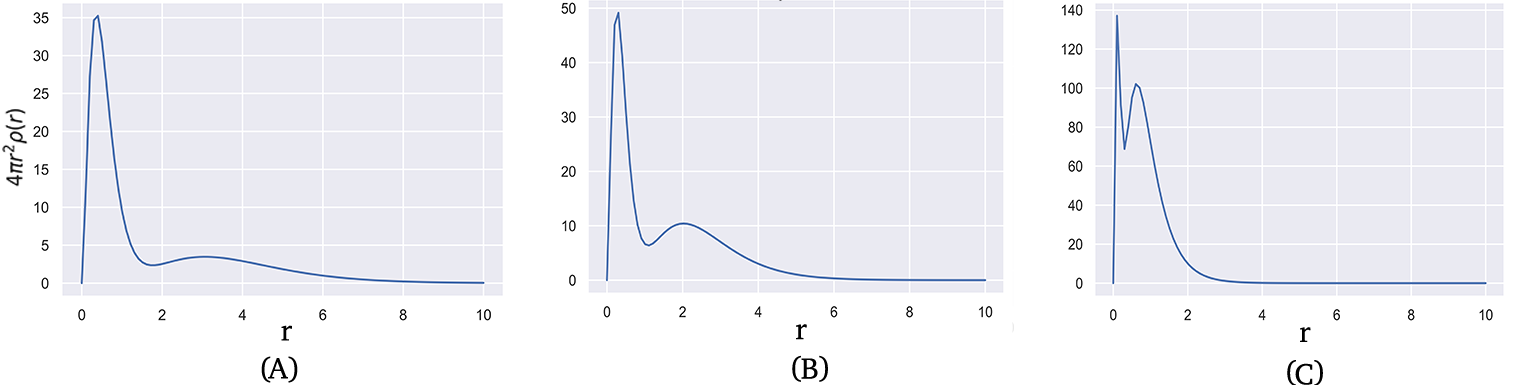}
\caption{Atomic probability densities for A) Lithium $\protect^2S$; B) Beryllium $^1S$; C) Neon $^1S$.}
\label{fig:All_Density}
\end{figure}

For this study the ground state atomic densities for Lithium $\protect^2S$ (Li),  Beryllium $^1S$ (Be), and  Neon $^1S$ (Ne) were computed with an average of 484 CI term expansion for the interacting and non--interacting spaces, and a Slater type orbital (STO) basis with harmonics in a  $0\leq l\leq 5$ range. All densities (fig.~\ref{fig:All_Density})  were defined  in terms of the density matrix using the Slater--Condon rules for monoelectronic operators and only the $LS$ eigenfunctions that correspond to the leading CSFs within a threshold contribution of 0.1 $\mu \mathrm{Ha}$ for each of the ground states were considered.

\section{Electron Correlation Potential Neural Network (eCPNN)}~\label{ecpnn}
The Hamiltonian operator $\hat{\rm H}$, as defined in the previous section, is often conceptualized as the sum of the kinetic ($\hat{\rm T}$) and  potential energy ($\hat{\rm V}$) operators.
An alternative, equally valid conceptualization of $\hat{\rm H}$ is  in terms of a fluctuation operator, ($\hat{\rm H}-\hat{\rm H}_0$), which contain all the information regarding the  instantaneous spatial correlations among all electrons, Thus, in atomic units (a.u.),  
\begin{equation}\label{eqn:Hamilt}
    \hat{\rm H}=\hat{\rm T}+\hat{\rm V}= \hat{\rm H}_0+  (\hat{\rm H}-\hat{\rm H}_0)= -\frac{\hbar^{2}}{2m}\frac{\partial^2}{\partial \Vec{r}^{2}}+\hat{\rm V}(\Vec{r})\equiv -\frac{\hbar^{2}}{2m}\nabla_{\Vec{r}}^{2}+\hat{\rm V}(\Vec{r}).
\end{equation}   
Generally, the fluctuation or interactive term  (or $\hat{\rm V}$) is unknown and difficult to compute~\cite{bunge2005cracking}, whereas the non-interacting term $\hat{\rm H}_0$ (or  $\hat{\rm T}$) may be inferred using the second derivatives of the correlated FCI probability amplitude  with respect to its spatial coordinates.

An alternative way to characterize quantum systems is presented by reformulating  quantum phenomena as a solution of inverse problems (i.e, approximating  an effective function containing  all the important physical constrains that generated the observed outcomes). 
In this context, the probability densities ($\rho$) of  multi-electronic systems can be used to define a proxy function ($|\phi|=\sqrt{\rho}$) to describe the quantum system. Although this proxy does not contain  the complete knowledge of all the electronic  interactions and correlations, its information is enough to define a  NN that can construct an effective Hamiltonian to accurately describe the quantum system under study. We propose to train our physics informed deep learning  model~\cite{raissi2017physics,raissi2017physicsII,raissi2019physics} via the following equation
\begin{equation}\label{eqn:FCI_loss}
    L_{EC}(\theta) = \bigg\lvert\bigg\rvert D_r \left(-\frac{\hbar^{2}}{2m} \frac{\nabla^2_{\Vec{r}} |\phi|}{|\phi|} + U_{\theta}(\Vec{r})\right)\bigg\lvert\bigg\rvert_{2}^2,
\end{equation}
where $D_r$ is the total derivative operator (with respect to $\Vec{r}$) acting on multi-variate function $-\frac{\hbar^{2}}{2m} \frac{\nabla^2_{\Vec{r}} |\phi|}{|\phi|} + U_{\theta}(\Vec{r})$, $\lvert\rvert \cdot \lvert\rvert_{2}$ is the Frobenius norm, and $U_{\theta}$ the learned parametric function that approximates the electronic correlation for each many--electron system. Because $U_{\theta}$ is given by a differential equation, the eCPNN is capable of approximating the effective potential $V(\Vec{r})$ for atomic systems up to an arbitrary constant. An initial condition is introduced to ensure the uniqueness of the solution via the Picard–Lindel\"of theorem. Equation~\ref{eqn:FCI_loss}, termed the effective correlation (EC) loss function, not only obeys the theoretical referents  of the FCI matrix-eigenvalue equation (Eq.~\ref{eq:2}), but effectively demands energy conservation for the many--electron quantum systems. Thus, for the proposed model, the complete loss function reads  %
   \begin{equation}\label{eqn:TISE_loss}
     L(\theta) = L_{EC}(\theta) + (U_{\theta}(\Vec{r}) - y)^2,
 \end{equation}
where $\Vec{r}$ is some point in the domain of the function and $y$ is the expected ground truth value for the true correlation function at that point. It is important to point out that  using  $|\phi|$ instead of  $\Psi^{FCI}$ for solving the electronic correlation problem may lead to the incorrect fluctuation potential at finitely many points where $\Psi^{FCI}$ changes signs. However, this last does not cause any difficulties in training our network. 

In our experiments, the eCPNN model is a $3$ layer feedforward network with hidden sizes of $64, 128$ and $128$ with $\tanh$ non--linearity within the hidden layers and a residual connection between the second and third layer. The inputs to the model are the $\Vec{r}$ spatial coordinates. For the network training, $5000$ of these coordinates were randomly selected from the domain of definition of each atomic system. The model was trained for $500$ epochs with Adam optimizer~\cite{kingma2017adam} and a learning rate of $0.001$. For stable training, we used a $L^2$ regularization on the weights of the network with the weight tuning parameter of $0.0001$.

\begin{figure}
\includegraphics[width=\linewidth,scale=0.90]{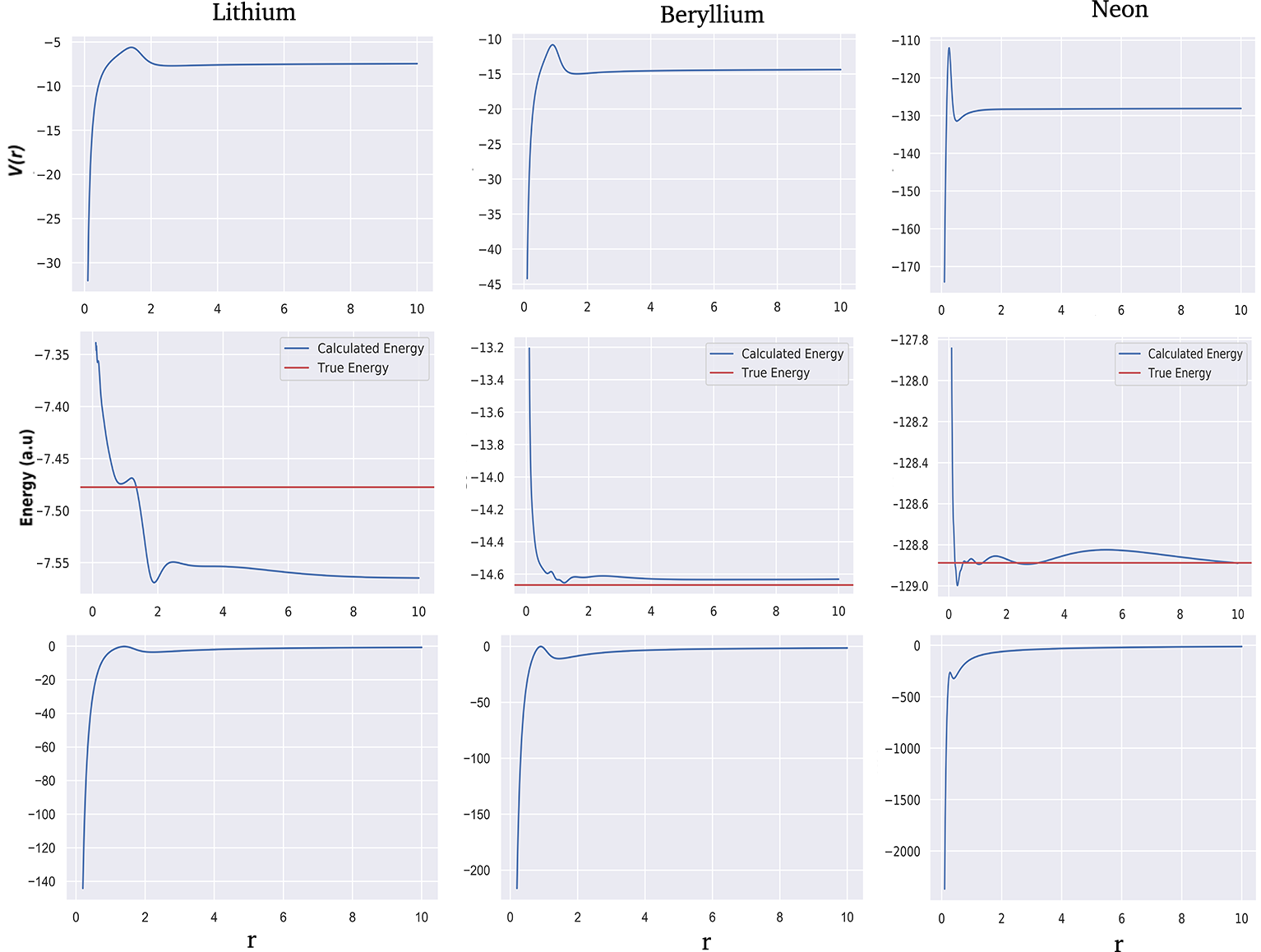}
\caption{ Learned correlation potentials (top), Learned energies (middle), virial theorem deviation (bottom).}
\label{fig:Potentials3}
\end{figure}
Our code is available at \url{https://github.com/arijitthegame/Quantum-Hamiltonians}. 

\section{Experiments}~\label{experiment}
The eCPNN model was validated on the Li, Be and Ne atoms. For all the atoms, spatial coordinates are defined on the interval $[0.1,10]$. As explained in equation~(\ref{eqn:TISE_loss}), an appropriate initial condition is added for each atom to ensure that an unique potential function is learned. The learned effective potential functions for  Li, Be and Ne atoms present a maximum 
whose height and width correlate  with the number of electrons in each atom (fig.~\ref{fig:Potentials3}). The larger the number of electrons the taller and more narrow the maximum in the potential curve.  
%
%
Using our learned potential function and the pre-computed density, we estimate the total energy of the system. Table~\ref{tab:Energies} shows the FCI energies ($E_{FCI}$) for each of the atoms studied and the energies approximated by the eCPNN model.
\begin{table}[h]
\begin{center}
\scalebox{0.85}{
\begin{tabular}{l c c c} 
 System & $E_{FCI}$  &$E_{eCPNN}$  & \begin{tabular}[c]{@{}c@{}} RMSE: True vs. Learned Energies \end{tabular} \\
 \midrule
 Lithium Atom & $-7.477487$& $-7.5431$ & $0.07 \pm {0.04}$\\
 Beryllium Atom &  $-14.66101$ & $-14.6142$ & $0.01 \pm{0.02}$ \\
Neon Atom &  $-128.888004$& $-128.8584$ & $0.01 \pm {0.009}$ \\
 
\bottomrule
\end{tabular}}
\vspace{1 mm}
\caption{FCI and eCPNN Learned Energies in a.u.: The 2$^{\text{nd}}$ column shows the average energy calculated by eCPNN over 5 runs.}
\label{tab:Energies}
\end{center}
\end{table}
For the  atoms studied,  the largest error in the eCPNN predicted energy correspond to Li. Unlike the Be and Ne atoms with closed--shell reference configurations for the ground state, $1s^22s^2$ and $1s^22s^22p^6$ respectively, the reference configuration for Li is $1s^22s^1$ which correspond to an open--shell configuration with a unpaired electron in a $s$ shell. It is well known that computing the energy of open--shell systems is more challenging due to their electronic structure characterized by having different number of $\alpha$ and $\beta$ electrons. Thus,  different considerations for the exchange and correlation electronic effects are needed. However, for the closed--shell atoms, the energies computed by  eCPNN are remarkably close to the true energies ($E^{FCI}$). In the learned  $V(\Vec{r})$,  the attractive regions at small $\Vec{r}$ present an upturn close to the origin. The sharp maximum in $V(\Vec{r})$ seems to be an  attribute of the potential where its position coincides with the minimum in $4\pi r^2\rho(\Vec{r})$ separating the peaks in the density from the atomic shells. These results suggest  that the peak in $V(\Vec{r})$ separates
spatially the parallel spin electrons in the different shells. 
For more details, see Appendix A.1.

In the context of density functional theory (DFT), the virial theorem is commonly used as a constraint to calculate atomic energies within the Kohn-Sham DFT formalism and it is frequently utilized as an indicator of the quality of the approximation in computational chemistry tasks
\cite{lowdin1959scaling,RosasCrawford,Baskerville2019,levy85,averil81,ghoshparr85}. As shown in fig.~\ref{fig:Potentials3}, for all the studied atoms, the effective potentials learned by the eCPNN  also satisfy the virial theorem, $2\langle T \rangle +\langle V\rangle =0$.
\section{Conclusion}~\label{conclusion}
We present a deep learning framework called eCPNN which was trained using high fidelity FCI data in an unsupervised manner. The eCPNN can approximate compact potential functions capable of describing the complex electric correlation effects important for describing atomic systems. This presents the possibility of defining succinct functions capable of predicting highly accurate energies for other many--electron atomic systems. Furthermore, we show that our learned potential functions satisfy important physical constraints like the virial theorem. Our work opens up interesting research directions in various problems in catalysis, rovibrational spectroscopy, photoelectron spectroscopy, and excited-state chemistry and enzymatic processes. An important drawback of our study is the dependency of the defined potentials on the calculated high fidelity densities. Studies are underway to create more robust systems when a measurable amount of noise is present.

\section{Broader Impact}
We envision this work to be beneficial to a broader community since we hope it will encourage researchers to use deep learning in trying solve various complicated differential equations. We are however limited by the curse of dimensionality as it will be significantly difficult to be able to run these experiments on a CPU.

Quantum mechanics has been one of the most successful models for describing the physical world. However, quantum mechanical systems are generally hard to solve and exact solutions only exist for simple systems. As such, by leveraging the power of neural networks we have aimed to improve the practical use of quantum mechanics and thus potentially contribute to the understanding of our world. One caveat is that our method provides only an approximation for the description of  quantum phenomena, and thus the possibility of incorrect predictions cannot be precluded.

\bibliographystyle{unsrt}
\bibliography{rsc}

\begin{thebibliography}{10}

\bibitem{Dral2021}
Pavlo~O. Dral, Fuchun Ge, Bao-Xin Xue, Yi-Fan Hou, Max Pinheiro, Jianxing
  Huang, and Mario Barbatti.
\newblock Mlatom 2: An integrative platform for atomistic machine learning.
\newblock {\em Top. Curr. Chem.}, 379(4):27, Jun 2021.

\bibitem{behler2011atom}
J{\"o}rg Behler.
\newblock Atom-centered symmetry functions for constructing high-dimensional
  neural network potentials.
\newblock {\em J. Chem. Phys.}, 134(7):074106, 2011.

\bibitem{QPNN}
Arijit Sehanobish, Hector~H. Corzo, Onur Kara, and David van Dijk.
\newblock Learning potentials of quantum systems using deep neural networks.
\newblock In Jonghyun Lee, Eric~F. Darve, Peter~K. Kitanidis, Michael~W.
  Mahoney, Anuj Karpatne, Matthew~W. Farthing, and Tyler Hesser, editors, {\em
  Proceedings of the AAAI 2021 Spring Symposium on Combining Artificial
  Intelligence and Machine Learning with Physical Sciences}, pages 1--13, 2021.

\bibitem{Manzhos2020}
Sergei Manzhos.
\newblock Machine learning for the solution of the schrödinger equation.
\newblock {\em Machine Learning: Science and Technology}, 1(1):013002, apr
  2020.

\bibitem{Zhang2018}
Linfeng Zhang, Jiequn Han, Han Wang, Roberto Car, and Weinan E.
\newblock Deep potential molecular dynamics: A scalable model with the accuracy
  of quantum mechanics.
\newblock {\em Phys. Rev. Lett.}, 120:143001, Apr 2018.

\bibitem{NoeSci}
Frank Noé, Simon Olsson, Jonas Köhler, and Hao Wu.
\newblock Boltzmann generators: Sampling equilibrium states of many-body
  systems with deep learning.
\newblock {\em Science}, 365(6457):1147, 2019.

\bibitem{yang2019molecular}
Qingyi Yang, Vishnu Sresht, Peter Bolgar, Xinjun Hou, Jacquelyn~L Klug-McLeod,
  Christopher~R Butler, et~al.
\newblock Molecular transformer unifies reaction prediction and retrosynthesis
  across pharma chemical space.
\newblock {\em Chem. Comm.}, 55(81):12152--12155, 2019.

\bibitem{CarleoSci}
Giuseppe Carleo and Matthias Troyer.
\newblock Solving the quantum many-body problem with artificial neural
  networks.
\newblock {\em Science}, 355(6325):602--606, 2017.

\bibitem{Pfau2020}
David Pfau, James~S. Spencer, Alexander G. D.~G. Matthews, and W.~M.~C.
  Foulkes.
\newblock Ab initio solution of the many-electron schr\"odinger equation with
  deep neural networks.
\newblock {\em Phys. Rev. Research}, 2:033429, Sep 2020.

\bibitem{ma2020deep}
Xinran Ma, ZC~Tu, and Shi-Ju Ran.
\newblock Deep neural network predicts parameters of quantum many-body
  hamiltonians by learning visualized wave-functions.
\newblock {\em arXiv preprint arXiv:2012.03019}, 2020.

\bibitem{lowdin1959scaling}
Per-Olov L{\"o}wdin.
\newblock Scaling problem, virial theorem, and connected relations in quantum
  mechanics.
\newblock {\em J. Mol. Spectrosc}, 3(1-6):46--66, 1959.

\bibitem{marc1985virial}
Guilhem Marc and WG~McMillan.
\newblock The virial theorem.
\newblock {\em Adv. Chem. Phys}, 58:209--361, 1985.

\bibitem{Fock1930}
V.~Fock.
\newblock Bemerkung zum virialsatz.
\newblock {\em Zeitschrift f{\"u}r Physik}, 63(11):855--858, Nov 1930.

\bibitem{shavitt1977method}
Isaiah Shavitt.
\newblock The method of configuration interaction.
\newblock In {\em Methods of electronic structure theory}, pages 189--275.
  Springer, 1977.

\bibitem{bunge2018present}
Carlos~F Bunge.
\newblock Present status of selected configuration interaction with truncation
  energy error.
\newblock {\em Adv. Quantum Chem.}, 76:3--34, 2018.

\bibitem{barr1970nature}
Tery~L Barr and Ernest~R Davidson.
\newblock Nature of the configuration-interaction method in ab initio
  calculations. {I}. {Ne} ground state.
\newblock {\em Phys. Rev. A}, 1(3):644, 1970.

\bibitem{bunge2005cracking}
Carlos~F Bunge.
\newblock Cracking electron correlation.
\newblock {\em Phys. Scr.}, 2005(T120):78, 2005.

\bibitem{bunge2006select}
Carlos~F Bunge and Ramon Carb{\'o}-Dorca.
\newblock Select-divide-and-conquer method for large-scale configuration
  interaction.
\newblock {\em J. Chem. Phys.}, 125(1):014108, 2006.

\bibitem{raissi2017physics}
Maziar Raissi, Paris Perdikaris, and George~Em Karniadakis.
\newblock {Physics informed deep learning (part I):} data-driven solutions of
  nonlinear partial differential equations.
\newblock {\em arXiv preprint arXiv:1711.10561}, 2017.

\bibitem{raissi2017physicsII}
Maziar Raissi, Paris Perdikaris, and George~Em Karniadakis.
\newblock {Physics Informed Deep Learning (Part II):} data-driven discovery of
  nonlinear partial differential equations.
\newblock {\em arXiv preprint arXiv:1711.10566}, 2017.

\bibitem{raissi2019physics}
Maziar Raissi, Paris Perdikaris, and George~E Karniadakis.
\newblock Physics-informed neural networks: A deep learning framework for
  solving forward and inverse problems involving nonlinear partial differential
  equations.
\newblock {\em J. Comput. Phys}, 378:686--707, 2019.

\bibitem{kingma2017adam}
Diederik~P. Kingma and Jimmy Ba.
\newblock Adam: A method for stochastic optimization.
\newblock {\em arXiv preprint arXiv:1412.6980}, 2017.

\bibitem{RosasCrawford}
Victor~M. Rosas-Garcia and T.~Daniel Crawford.
\newblock The electron cusp condition and the virial ratio as indicators of
  basis set quality.
\newblock {\em J. Chem. Phys.}, 118(6):2491--2497, 2003.

\bibitem{Baskerville2019}
Adam~L. Baskerville, Andrew~W. King, and Hazel Cox.
\newblock Electron correlation in {Li$^+$}, {He}, {H$^-$} and the critical
  nuclear charge system z$_c$: energies, densities and coulomb holes.
\newblock {\em R. Soc. Open Sci.}, 6(1):181357, 2019.

\bibitem{levy85}
Mel Levy and John~P. Perdew.
\newblock Hellmann-feynman, virial, and scaling requisites for the exact
  universal density functionals. shape of the correlation potential and
  diamagnetic susceptibility for atoms.
\newblock {\em Phys. Rev. A}, 32:2010--2021, Oct 1985.

\bibitem{averil81}
F.~W. Averill and G.~S. Painter.
\newblock Virial theorem in the density-functional formalism: Forces in
  {${\mathrm{H}}_{2}$}.
\newblock {\em Phys. Rev. B}, 24:6795--6800, Dec 1981.

\bibitem{ghoshparr85}
Swapan~K. Ghosh and Robert~G. Parr.
\newblock Density‐determined orthonormal orbital approach to atomic energy
  functionalsa).
\newblock {\em J. Chem. Phys.}, 82(7):3307--3315, 1985.

\bibitem{hoggan2013proceedings}
P.E. Hoggan.
\newblock {\em Proceedings of MEST 2012: Exponential Type Orbitals for
  Molecular Electronic Structure Theory}.
\newblock ISSN. Elsevier Science, 2013.

\end{thebibliography}
\newpage
\appendix
\section{Appendix}
\subsection{Electron Density}
\begin{figure}[h]
	\centering
	\includegraphics[width=0.40\textwidth]{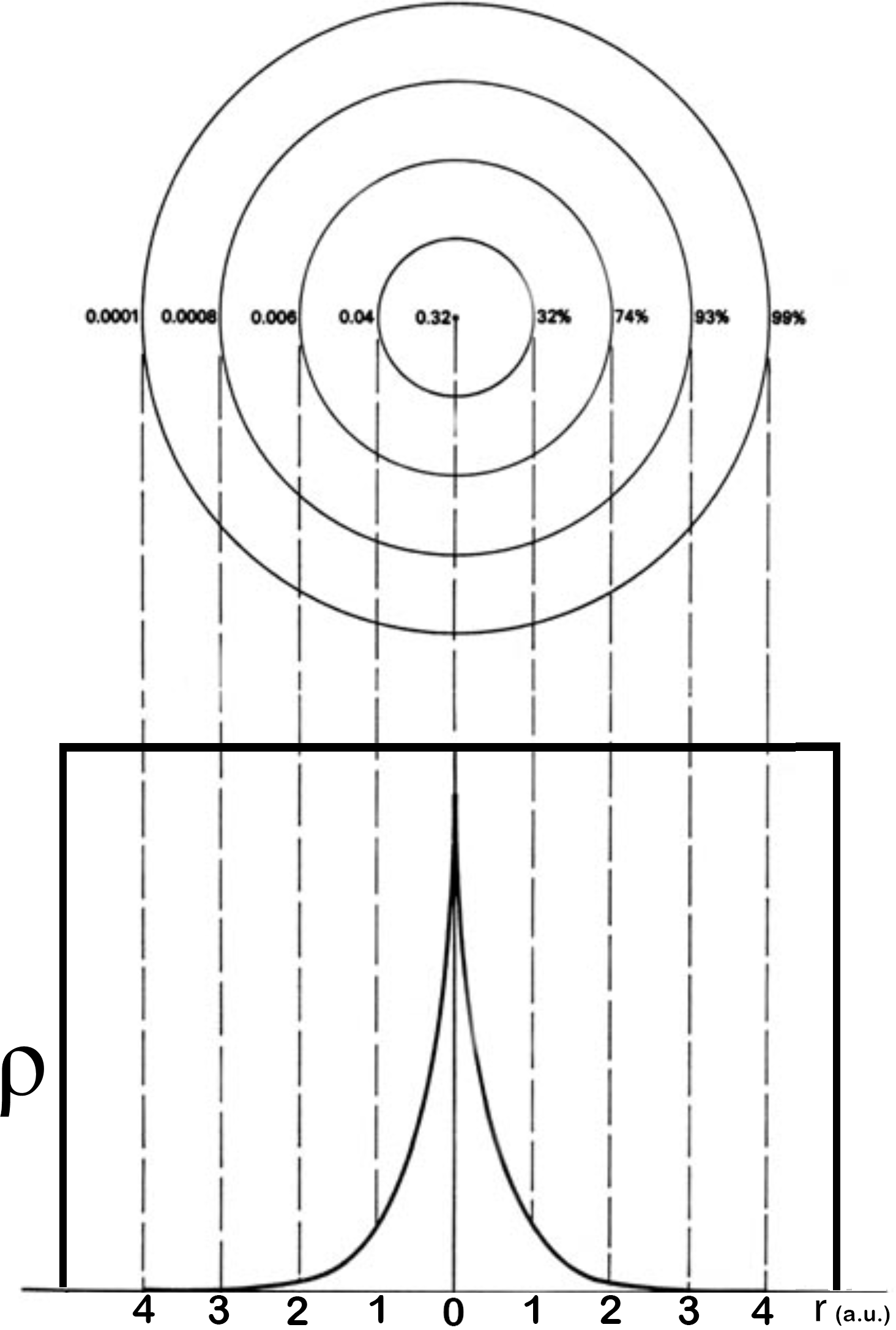}
	\caption{Cusp behavior of $\rho$ near r=0}
	\label{Fig: Cusp}
\end{figure}

Mathematically, the function $\rho(r)$ exhibits non-analytical behaviour in the form of a discontinuity of its gradient exactly at the origin r = 0, (i.e. the positions of the nuclei) which results poles in the potential energy function at these positions (V = $-\infty$)\cite{hoggan2013proceedings}. The electronic wave--function must satisfy the cusp condition (which result in spikes of $\rho(r)$ here) in the neighbourhood of each of the nuclei electronic wave--function, where $\rho(r)\sim e^{-2Zr}$. Note: How sharp the spikes observed around the cusp depends on the charge of the nucleus Z such that an infinitesimal deviation from the position of the nucleus has to be accompanied by such a decreasing of the density given by $\lim _{r \rightarrow 0} \frac{\mathrm{d} \ln \rho(r)}{\mathrm{d} r}=-2 Z$.
Figure \ref{Fig: Cusp} illustrates a contour map of the electron density distribution in a plane containing the nucleus for the n = 1 level of the H atom. The distance between adjacent contours is 1 atomic unit (a.u.). The numbers on the left-hand side on each contour correspond to the electron density in a.u. The numbers on the right-hand side correspond to the proportion of the total electronic charge  which lies within a sphere of that radius. The figure therefore reflects the fact that 99\% of the single electronic charge of the H atom lies within a sphere of radius 4 a.u.

\subsection{Local potential}

Defining an eigenvalue equation for finding the local potential $V(r)$ reads  

\begin{equation}
    [\frac{1}{2}\nabla_{\Vec{r}}^{2} + V(r)]\phi=\varepsilon\phi,~\rm{(a.u.)}
\end{equation}

where the electronic density is just
\begin{equation}
    \rho=2|\phi|^2.
\end{equation}

Thus,

\begin{equation}
    \phi=(\frac{1}{2}\rho)^{1/2}
\end{equation}
and $V(r)$ may be computed from 
\begin{equation}
   V(r)=\phi^{-1}(\varepsilon + \frac{1}{2}\nabla^{2})\phi
\end{equation}
Since the exact $\rho$ is positive and monotonic-decreasing as $r$ increases, it can be chosen to be positive everywhere without introducing discontinuities in  its derivatives. Conventionally, $\varepsilon$ is chosen so that $V(r)\rightarrow 0$ as $r\rightarrow \infty$. For the exact $\rho$, this requires $\varepsilon = -I$, where $I$ is the ionization energy of the atomic or molecular system under study.

\newpage
\end{document}